\documentclass[aps,prl,twocolumn,superscriptaddress]{revtex4}
\usepackage{epsf,graphicx}
\usepackage{amssymb}
\usepackage{amsmath}
\usepackage{bm}
\usepackage{epstopdf}
\usepackage{color}
\usepackage{ulem}
\begin{document}
\title{Biorthogonal Dynamical Quantum Phase Transitions in Non-Hermitian Systems}
\author{Yecheng Jing}
\affiliation{Department of Physics and Chongqing Key Laboratory for Strongly Coupled Physics, Chongqing University, Chongqing 401331, China}
\author{Jian-Jun Dong}
\email[]{dongjianjun@cqu.edu.cn}
\affiliation{Department of Physics and Chongqing Key Laboratory for Strongly Coupled Physics, Chongqing University, Chongqing 401331, China}
\author{Yu-Yu Zhang}
\email[]{yuyuzh@cqu.edu.cn}
\affiliation{Department of Physics and Chongqing Key Laboratory for Strongly Coupled Physics, Chongqing University, Chongqing 401331, China}
\author{Zi-Xiang Hu}
\email[]{zxhu@cqu.edu.cn}
\affiliation{Department of Physics and Chongqing Key Laboratory for Strongly Coupled Physics, Chongqing University, Chongqing 401331, China}
\date{\today}

\begin{abstract}
By utilizing biorthogonal bases, we develop a comprehensive framework for studying biorthogonal dynamical quantum phase transitions in non-Hermitian systems. With the help of the previously overlooked associated state, we define the automatically normalized biorthogonal Loschmidt echo. This approach is capable of handling arbitrary non-Hermitian systems with complex eigenvalues and naturally eliminates the negative value of Loschmidt rate obtained without the biorthogonal bases. Taking the non-Hermitian Su-Schrieffer-Heeger model as a concrete example, a $1/2$ change of dynamical topological order parameter in biorthogonal bases is observed which is not shown in self-normal bases. Furthermore, we discover that the periodicity of biorthogonal dynamical quantum phase transitions depends on whether the two-level subsystem at the critical momentum oscillates or reaches a steady state.
\end{abstract}

\maketitle

The past decades have witnessed the flourishing of non-Hermitian physics in nonconservative systems as found in a variety of physical realms including open quantum systems \cite{Rotter2009JPA}, electronic systems with interactions \cite{Yoshida2018PRB,Shen2018PRL,Fu2020PRL}, and classical systems with gain or loss \cite{Makris2008PRL,Klaiman2008PRL,Malzard2015PRL}. In these systems \cite{Bender1998PRL,Helbig2020NatPhys,Liu2021Research,Weidemann2020Science,Brandenbourger2019NatCommun,
Ghatak2020PNAS,XZhang2021NatCommun,LZhang2021NatCommun,Xiao2020NatPhys,Wang2021JOpt,ZhongWang2018PRL,Sato2019PRX,Yokomizo2019PRL,Shen2018PRL2,Balabanov2021PRR,Guo2023PRL}, many novel physics and unprecedented phenomena have been explored recently, such as the exceptional points \cite{Hodaei2017Nature,Bergholtz2021RMP}, the non-Hermitian skin effects \cite{Zhang2020PRL,XiujuanZhang2022AdvPX}, the bulk Fermi arcs \cite{Zhou2018Science}, and so on. In contrast to the Hermitian systems with real eigenvalues and orthogonal eigenstates, the eigenvalues and eigenstates in a general non-Hermitian Hamiltonian are not necessarily real and orthogonal \cite{Ueda2020AdvPhys}. To be more precise, the orthogonality of eigenstates is replaced by the notion of biorthogonality that defines the relation between the Hilbert space of states and its dual space, leading to the so-called ``biorthogonal quantum mechanics" \cite{Brody2014JPA,Emil2018PRL,Emil2019PRBR,Emil2020PRR}. A direct consequence of the biorthogonality is that the transition probability between one state $\left\vert \phi\right\rangle$ to its time-evolved state $\left\vert \phi \left(t \right) \right\rangle$ should be carefully defined. The traditional viewpoint $p=\left\vert \left\langle \phi \left(t \right)|\phi\right\rangle \right\vert ^{2}$ used in Hermitian systems cannot be applied to a general non-Hermitian Hamiltonian. A schematic framework to deal with this becomes an urgent topic due to the rapid development of nonequilibrium studys in non-Hermitian systems \cite{ShuChen2021PRA,HuiZhai2020NatPhys,Lin2022NPJQI,Liu2023PRA,Hauke2022PRXQ,Kawabata2023PRX,Yoshimura2020PRB,Mcdonald2022PRB,Yin2022PRB,Agarwal2022arXiv,Agarwal2023arXiv,Roubeas2023JHEP}.

Dynamical quantum phase transition (DQPT) is arguably one of the most important nonequilibrium phenomena in modern many-body physics and has also been extensively studied in past decades \cite{Heyl2013PRL,Budich2016PRB,Dong2019PRB,Nie2020PRL,ShuChen2023PRB}. It was first introduced in the Hermitian transverse field Ising model \cite{Heyl2013PRL} and was generalized to mixed state \cite{Heyl2017PRB,Bhattacharya2017PRB}, finite temperature \cite{Abeling2016PRB,Sedlmayr2018PRB,Halimeh2018PRL,Halimeh2018PRB,Mera2018PRB}, Floquet systems \cite{Yang2019PRB,Naji2022PRA,Jafari2021PRA,Jafari2022PRB}, and slow quench process \cite{Sharma2016PRB}. It was also observed experimentally with trapped ions \cite{Shen2017PRL,Zhang2017Nature}, Rydberg atoms \cite{Bernien2017Nature}, ultracold atoms \cite{Sengstock2018NatPhys}, superconducting qubits \cite{HengFan2019PRAppl}, nanomechanical and photonic systems \cite{JiangfengDu2019PRB,PengXue2019PRL,GuangcanGuo2020LSA}. The key quantity characterizing DQPT is the Loschmidt echo or dynamical fidelity, $\mathcal{L}\left(  t\right)  \equiv\left\vert \left\langle \Psi\left( 0\right)  |\Psi\left(  t\right)  \right\rangle \right\vert ^{2}$, quantifying the time-dependent deviation from an arbitrary initial state $\left\vert \Psi\left(  0\right)  \right\rangle$. In Hermitian systems, DQPTs occur whenever the time-evolved state $\left\vert \Psi\left(  t\right)  \right\rangle$ becomes orthogonal to the initial state $\left\vert \Psi\left(  0\right)  \right\rangle$ and the critical time $t_c$ is defined as $\mathcal{L}\left(  t_c \right) =0$. Efforts have been made to generalize this concept to non-Hermitian systems, leading to many interesting predictions such as the half-integer jumps in dynamical topological order parameter (DTOP) \cite{Mondal2022PRB}, but the special biorthogonality has been ignored and an enforced normalized factor has been used \cite{Mondal2022PRB,Zhou2018PRA,Zhou2021NJP,Mondal2022arxiv}. Very recently, it has been shown that the non-Hermitian systems should be described by the biorthogonal fidelity and the biorthogonal Loschmidt echo instead of the conventional counterparts in Hermitian systems \cite{ShuChen2018PRA,Huang2021PRR}, but it is limited to the parity-time symmetry cases \cite{Sun2022FP,Tang2022EPL}. A natural treatment for a general non-Hermitian Hamiltonian is still lacking, which severely limits our exploration of the richness of DQPTs in these systems.

In this Letter, we address this issue by proposing a new theoretical framework to deal with these nonequilibrium phenomena in general non-Hermitian systems. Based on biorthogonal quantum mechanics, we reformulate the transition probability between $\left\vert \Psi\left(  0\right)  \right\rangle$ and $\left\vert \Psi\left(  t\right)  \right\rangle$ with the biorthogonal bases and the associated states. We introduce the concept of biorthogonal dynamical quantum phase transitions and compare it with its self-normal counterpart, utilizing the non-Hermitian Su-Schrieffer-Heeger model as a concrete example. We conduct a comprehensive investigation of the biorthogonal Loschmidt rate, biorthogonal DTOP, Fisher zeros, and the transition probability in momentum space during a sudden quench. Our calculations reveal a half-integer jump in the biorthogonal DTOP and demonstrate that the periodicity of biorthogonal DQPTs depends on whether the two-level subsystem at the critical momentum undergoes oscillation or settles into a steady state. Our theory establishes a general framework for studying nonequilibrium DQPTs in non-Hermitian systems.

We first review some basic properties of biorthogonal quantum mechanics in non-Hermitian systems and reformulate the probability assignment rules between two states with biorthogonal bases \cite{Brody2014JPA}. For a general non-Hermitian Hamiltonian $H\neq H^{\dag}$, the eigenvalue equations of $H$ and $H^{\dag}$ are given by
\begin{align}
H|u_{n}\rangle &  =\epsilon_{n}|u_{n}\rangle,\quad \langle u_{n}|H^{\dagger}=\epsilon_{n}^{\ast}\langle u_{n}|,\nonumber\\
H^{\dag}\left\vert \widetilde{u}_{n}\right\rangle  &  =\epsilon_{n}^{\ast}\left\vert \widetilde{u}_{n}\right\rangle ,\quad\left\langle \widetilde{u}_{n}\right\vert H=\epsilon_{n}\left\langle \widetilde{u}_{n}\right\vert, \label{eigenvalue-equations}%
\end{align}
where $\epsilon_{n}$ is the \textit{n}th eigenvalue, and $|u_{n}\rangle$ and $\left\vert \widetilde{u}_{n}\right\rangle $ are the right and left eigenstates that satisfy the completeness relation $\sum_{n} |\widetilde{u}_{n}\rangle\langle u_{n}|=1$ and the biorthonormal relation $\langle\widetilde{u}_{m}|u_{n}\rangle=\delta_{m,n}$. Note that under this condition $\langle\widetilde{u}_{n}|u_{n}\rangle=1$, eigenstates are no longer normalized. In particular, we have $\langle u_{n}|u_{n}\rangle\geq1$ and $\langle u_{n}|u_{m}\rangle\neq0$ if $n\neq m$. It challenges the traditional probabilistic interpretation used in Hermitian quantum mechanics. For instance, there cannot be a ``transition'' from one eigenstate $|u_{m}\rangle$ to another eigenstate $|u_{n}\rangle$ due to the orthonormal relation $\langle u_{n}|u_{m}\rangle=0$ if $n\neq m$ in Hermitian systems. To reconcile these apparent contradictions we need the introduction of the so-called associated state and the redefinition of the inner product.
For an arbitrary state $|\psi\rangle$, its associated state $|\widetilde{\psi}\rangle$ is defined according to the following relation \cite{Brody2014JPA}:
\begin{equation}
|\psi\rangle=\sum_{n}c_{n}|u_{n}\rangle\longleftrightarrow|\widetilde{\psi}\rangle=\sum_{n}c_{n}|\widetilde{u}_{n}\rangle, \label{associated-state}
\end{equation}
while the dual state $\langle \widetilde{\psi}\vert =\sum_{n}c_{n}^{\ast}\left\langle\widetilde{u}_{n}\right\vert$ is given by the Hermitian conjugate of $\vert \widetilde{\psi} \rangle$. The inner product between $|\psi\rangle$ and another state $|\phi\rangle=\sum_{m}d_{m}|u_{m}\rangle$ is thus defined as
\begin{equation}
\langle\phi,\psi\rangle\equiv\langle\widetilde{\phi}|\psi\rangle=\sum_{m,n}\langle\widetilde{u}_{m}|d_{m}^{\ast}c_{n}|u_{n}\rangle=\sum_{n}d_{n}^{\ast}c_{n}.
\end{equation}
It is easy to show that the norm of a state $|\psi\rangle$ is $\sqrt{\langle\widetilde{\psi}|\psi\rangle}$. With these new definitions, the transition probability between $|\psi\rangle$ and $|\phi\rangle$ for a biorthogonal system is given by
\begin{equation}
p=\frac{\langle\widetilde{\psi}|\phi\rangle\langle\widetilde{\phi}|\psi \rangle}{\langle\widetilde{\psi}|\psi\rangle\langle\widetilde{\phi} |\phi\rangle}, \label{probability}
\end{equation}
where the denominator acts as a natural normalizing factor. In this case, $p$ is a real number ranging from $0$ to $1$, which meets the requirements of probability interpretation satisfactorily. If the Hamiltonian is Hermitian $H= H^{\dag}$, we have $|\widetilde{u}_{n}\rangle=|u_{n}\rangle$ and $|\widetilde{\psi}\rangle=|\psi\rangle$. Then Eq.~\eqref{probability} reduces to the conventional definition of the transition probability in Hermitian quantum mechanics $p=\langle\psi|\phi\rangle\langle\phi|\psi\rangle/\left(  \langle\psi |\psi\rangle\langle\phi|\phi\rangle\right)$. Another important consequence of Eq.~\eqref{probability} is that the projection from $|\psi\rangle=\sum_{n}c_{n}|u_{n}\rangle$ to $|u_{n}\rangle$ becomes
\begin{equation}
p_{n}=\frac{\langle\widetilde{\psi}|u_{n}\rangle\langle\widetilde{u}_{n} |\psi\rangle}{\langle\widetilde{\psi}|\psi\rangle\langle\widetilde{u}_{n}|u_{n}\rangle}=\frac{c_{n}^{\ast}c_{n}}{\sum_{m}c_{m}^{\ast}c_{m}}, \label{projection}
\end{equation}%
satisfying the normalization condition $\sum_{n}p_{n}=1$.

By now the static aspects of non-Hermitian systems are clear, but the dynamical problems remain controversial. Although the time evolution of an arbitrary initial state $|\Psi\left(  0\right)  \rangle$ under a time-independent Hamiltonian $H$ is given by the well-known Schr\"{o}dinger equation $|\Psi\left(  t\right)  \rangle=\operatorname{e}^{-\operatorname*{i} Ht}\left\vert \Psi\left(  0\right)  \right\rangle$, the direct generalization $|\widetilde{\Psi}\left(  t\right)  \rangle=\operatorname{e}^{-\operatorname*{i}H^{\dag}t}|\widetilde{\Psi}\left(  0\right)  \rangle$ \cite{Tang2022EPL} may lead to unreasonable complex probabilities if there is no parity-time symmetry \cite{SM}. To establish a general framework for non-Hermitian Hamiltonian with complex eigenvalues, we use the associated state $|\widetilde{\Psi}\left(  t\right)  \rangle=\sum_{n}c_{n}|\widetilde{u}_{n}\rangle$ where $c_{n}=\langle\widetilde{u}_{n} |\Psi\left(  t\right)\rangle$ to deal with the dynamical problem. The overlap between $|\Psi\left(  0\right)  \rangle$ and $|\Psi\left(  t\right)  \rangle$ is characterized by the biorthogonal Loschmidt echo,
\begin{equation}
\mathcal{L}\left(  t\right)  =\frac{\langle\widetilde{\Psi}\left(  0\right)|\Psi\left(  t\right)  \rangle\langle\widetilde{\Psi}\left(  t\right)|\Psi\left(  0\right)  \rangle}{\langle\widetilde{\Psi}\left(  t\right) |\Psi\left(  t\right)  \rangle\langle\widetilde{\Psi}\left(  0\right) |\Psi\left(  0\right)  \rangle}.
\end{equation}%
As in the Hermitian case, the biorthogonal DQPTs can be defined as $\mathcal{L}\left(  t_c \right) =0$ with the critical time $t_c$.

We may examine the biorthogonal DQPT by considering a general non-Hermitian Hamiltonian $H=\sum_{k}\psi_{k}^{\dag}H_{k}\psi_{k}$ with $ H_{k}=\bm{d}_{k}\cdot\bm{\sigma}$. Here, $\bm{\sigma}$ is the vector of the Pauli matrices, $\bm{d}_{k}=\left( x_{k},y_{k},z_{k}\right)$ represents the vector of expansion coefficients, and at least one component of $\bm{d}_{k}$ is complex in non-Hermitian systems. The eigenenergies of $H_k$ are given by $\pm\epsilon_k=\pm\sqrt{\bm{d}_k^2}$, and the corresponding eigenstates are $|u_{k\pm}\rangle$. Considering a quench process where the model Hamiltonian changes from $\bm{d}_k^{i}$ at time $t=0^-$ to $\bm{d}_k^{f}$ at time $t=0^+$, the initial state $|\Psi\left(  0\right)  \rangle={\textstyle\bigotimes\nolimits_{k}}|u_{k-}^{i}\rangle$ which is defined as the tensor product of all $|u_{k-}^i\rangle$ is evolved under the postquench Hamiltonian $\bm{d}_k^{f}$. The biorthogonal Loschmidt echo can be expressed as $\mathcal{L}(t)=\prod_{k}g_{k}(t)$ with \cite{SM}
\begin{align}
g_{k}(t)=\frac{|\cos(\epsilon_{k}^ft)-\operatorname*{i}\sin(\epsilon_{k}^ft)\langle
\widetilde{u}_{k-}^i|\frac{H_{k}^f}{\epsilon_{k}^f}|u_{k-}^i\rangle|^2}{\langle\widetilde{u}_{k-}^i(t)|u_{k-}^i(t)\rangle},
\label{gkt}
\end{align}
where $|u_{k-}^i(t)\rangle=\text{e}^{-\operatorname*{i}H^f_{k}t}|u_{k-}^i\rangle$. To obtain a nonzero and well-defined quantity in the thermodynamic limit it is useful to consider the biorthogonal Loschmidt rate
\begin{align}
\text{LR}(t)&=-\displaystyle\lim_{N\to\infty}\frac{1}{N}\ln{\mathcal{L}(t)},\label{LR}
\end{align}
where $N$ is the system size. Zeros in $\mathcal{L}(t)$ at critical times $t_c$ correspond to nonanalyticities (cusps or divergencies) in $\text{LR}(t)$. If there is at least one pair of critical parameters $k_c$ and $t_c$ such that $g_{k_c}(t_c)=0$, then $\mathcal{L}(t_c)=0$. The solution of $g_{k_c}(t_c)=0$ is
\begin{align}
t_c=\frac{\pi}{2\epsilon^f_{k_c}}(2n+1)-\frac{\operatorname*{i}}{\epsilon^f_{k_c}}\tanh^{-1}\langle \widetilde{u}_{k_c-}^i|\frac{H_{k_c}^f}{\epsilon_{k_c}^f}|u_{{k}_c-}^i\rangle,
\label{critical time}
\end{align}
where $n$ is an integer. If we can obtain a positive real solution $t_c$, the system will undergo a biorthogonal DQPT. In general, it is difficult to access $k_c$ in a finite-size system because momentum takes quantized values. Thus the divergence of $\text{LR}(t)$ in the thermodynamic limit becomes a cusp in a finite-size system except for some fine-tuned quench parameters or twist boundary conditions \cite{ShuChen2023PRB}. For the same reason, it is also difficult to obtain a positive real solution $t_c$ in a finite system. To address this issue, we can solve the equation $\mathcal{L}(t)=0$ inversely and study the distribution of the roots in the complex time plane. The solution of $\mathcal{L}(t_c)=0$ depends on the system size $N$. If for all $\epsilon >0$, there exists a natural number $N_0$ such that $\operatorname{Im} [t_c(N)]< \epsilon$ holds for all $N>N_0$, a real solution $\lim_{N\rightarrow\infty}\operatorname{Im}\left[  t_{c}\left(  N\right)\right]  =0$ can be obtained, indicating a biorthogonal DQPT.

\begin{figure}[ptb]
\begin{center}
\includegraphics[width=8.6cm]{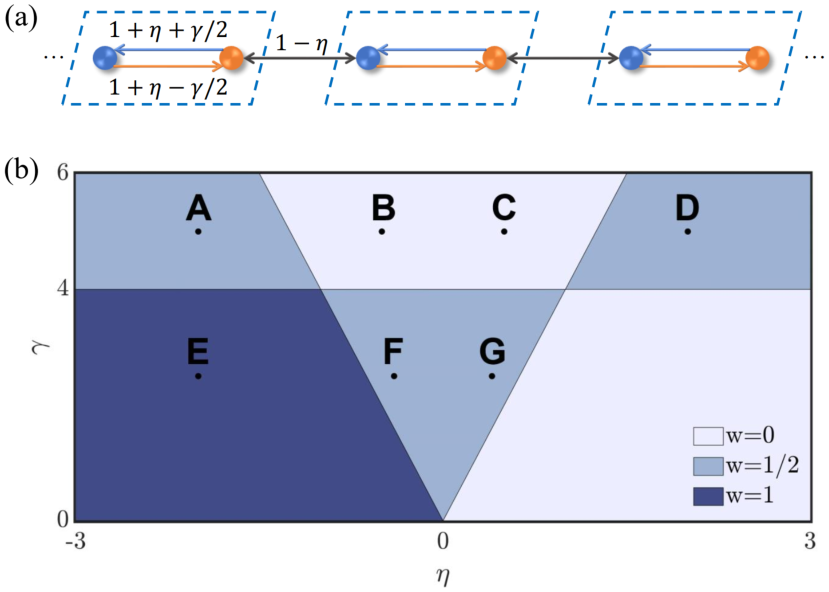}
\end{center}
\caption{(a) The sketch of non-Hermitian Su-Schrieffer-Heeger model. The box indicates the unit cell. (b) The $\gamma$-$\eta$ phase diagram. Each region filled with a different color corresponds to a phase with a topological winding number $w$. The boundaries between the regions are determined using $\epsilon_{k}=0$. The black points represent different parameters ($\eta$, $\gamma$) which will be used below. Specifically, the points A to G represent ($-2$, $5$), ($-0.2$, $5$), ($0.2$, $5$), ($2$, $5$), ($-2$, $1$), ($-0.2$, $1$), and ($0.2$, $1$), respectively.}
\label{fig1}
\end{figure}

Furthermore, similar to the DTOP in Hermitian systems \cite{Vajna2015PRB}, we can introduce a biorthogonal DTOP to describe biorthogonal DQPT. The biorthogonal DTOP is defined as
\begin{align}
\nu(t)=\frac{1}{2\pi}\int_{0}^{2\pi}dk\partial_k\phi_k^G(t),
\label{DTOP}
\end{align}
where the biorthogonal geometrical phase is $\phi^G_k(t)=\phi_k(t)-\phi_k^{\text{dyn}}(t)$, with $\phi_k(t)$ being the phase of $g_k(t)$ and the biorthogonal dynamical phase is given by \cite{SM}
\begin{align}	\phi_k^{\text{dyn}}(t)=&-\int_{0}^{t}ds\frac{\langle\widetilde{u}_{k-}^i(s)|H_k^f|u_{k-}^i(s)\rangle}{\langle\widetilde{u}_{k-}^i(s)|u_{k-}^i(s)\rangle}\nonumber\\
&+\frac{\text{i}}{2}\text{ln}\langle\widetilde{u}_{k-}^i(t)|u_{k-}^i(t)\rangle.  \label{dyn}
\end{align}

To demonstrate the efficacy of our novel theoretical framework in dealing with non-Hermitian Hamiltonians featuring complex eigenvalues, we present a comprehensive examination of the non-Hermitian Su-Schrieffer-Heeger model as outlined below. The Hamiltonian is
\begin{align}
H  & =\sum_{j}\left[  \left(  1+\eta+\frac{\gamma}{2}\right)  c_{j,b}^{\dag}c_{j,a}+\left(  1+\eta-\frac{\gamma}{2}\right)  c_{j,a}^{\dag}c_{j,b}\right.
\nonumber\\
& \left.  \quad+\left(  1-\eta\right)  c_{j,a}^{\dag}c_{j+1,b}+\left(1-\eta\right)  c_{j+1,b}^{\dag}c_{j,a}\right]  ,
\end{align}
where $\eta$ determines the strength of intracell and intercell hopping and $\gamma$ controls the degree of non-Hermiticity, as shown in Fig.~\ref{fig1}(a). For the periodic boundary condition, the bulk Hamiltonian gets the standard bilinear form $H=\sum_k\psi_k^{\dagger}\left(  \bm{d}_{k}\cdot\bm{\sigma}\right)\psi_k$ in momentum space, where $\psi_k^\dagger=(c_{k,a}^{\dag},c_{k,b}^{\dag})$ and
\begin{equation}
\bm{d}_{k}=\left(  \left(  1+\eta\right)  +\left(  1-\eta\right)  \cos k,\left(
1-\eta\right)  \sin k-\operatorname*{i}\frac{\gamma}{2},0\right).
\end{equation}
The dispersion is $\pm\epsilon_{k}=\pm\sqrt{x_{k}^{2}+y_{k}^{2}}$. It becomes gapless at the exceptional points. Thus the solution of $\epsilon_{k}=0$ determines the phase boundary, i.e., $k_c=0, \; \gamma=\pm4$ and $k_c=\pi,\; \gamma=\pm4\eta$. Combining with the winding number $w=\frac{1}{2\pi}\int_{0}^{2\pi}dk\partial_{k}\left(  \arctan \frac{y_{k}}{x_{k}}\right)$ \cite{Yin2018PRA}, we present the phase diagram in Fig.~\ref{fig1}(b) for convenience, where we only consider the case $\gamma>0$ due to the symmetry of the phase diagram with respect to $\gamma$.

\begin{figure}[ptb]
\begin{center}
\includegraphics[width=8.6cm]{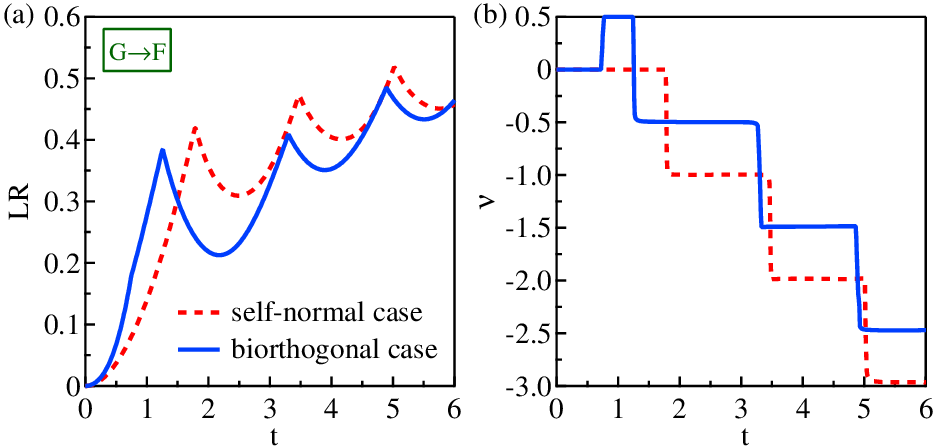}
\end{center}
\caption{(a) The Loschmidt rate and (b) the DTOP with self-normal and biorthogonal bases for a typical quench process (from G to F).}
\label{fig2}
\end{figure}

For comparison, we also calculate the traditional DQPT based on self-normal Loschmidt echo $\mathcal{L}\left(  t\right)  =\left\vert \left\langle \Psi\left( 0\right)  |\Psi\left(  t\right)  \right\rangle \right\vert ^{2}$, with an enforced normalized factor to avoid the negative value of Loschmidt rate \cite{Zhou2018PRA}. Figure~{\ref{fig2}}(a) presents the self-normal and biorthogonal Loschmidt rate from the same quenching process. The first feature is that the critical time at which the cusp appears is different from each other. A similar phenomenon has been observed in equilibrium quantum phase transitions, where the biorthogonal and self-normal fidelity predict different quantum critical points \cite{Sun2022FP}. It turns out that the biorthogonal fidelity captures the correct critical point due to the special biorthoganality in non-Hermitian systems \cite{Sun2022FP}. Similarly, the nonequilibrium quantum phase transitions should be described by the biorthogonal time-dependent
version of the fidelity, i.e., biorthogonal Loschmidt echo. The second feature is that there is an additional critical time $t_c\approx 0.74$ in biorthogonal bases. This can be seen more clearly in the DTOP as shown in Fig.~{\ref{fig2}}(b). A half-integer jump of $\nu(t)$ can be observed in the biorthogonal bases while there is no such change in the self-normal bases.

\begin{figure}[ptb]
\begin{center}
\includegraphics[width=8.6cm]{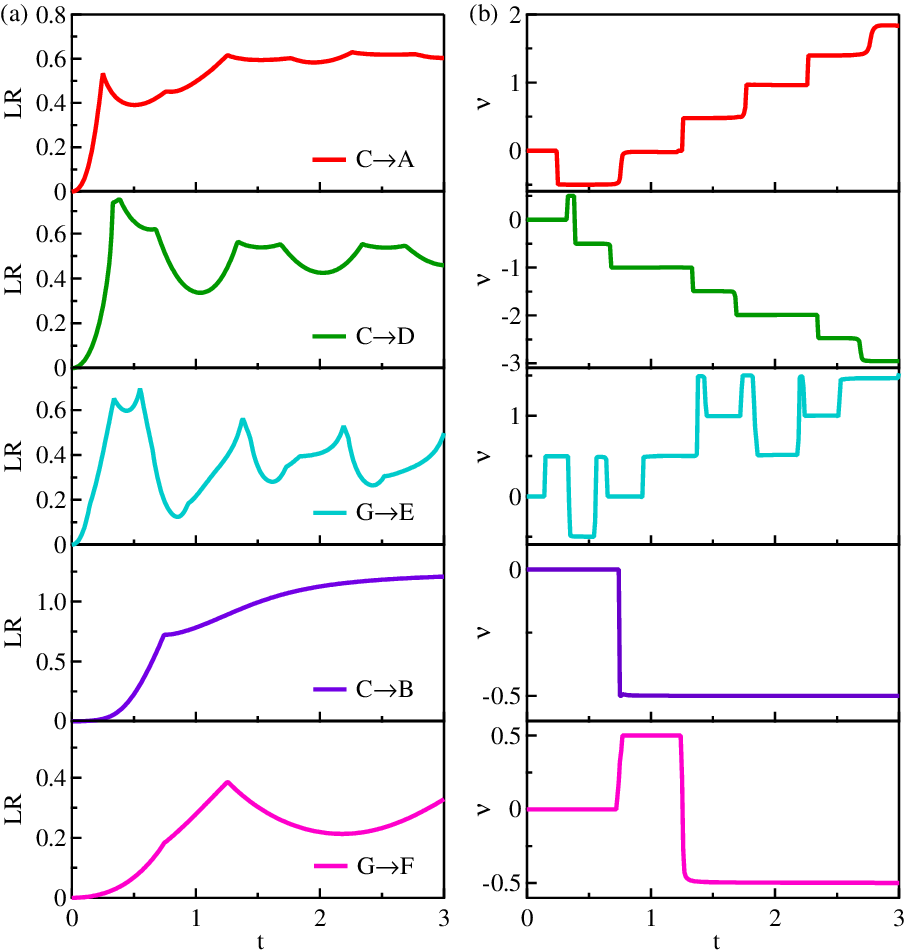}
\end{center}
\caption{Five different types of quenching processes showing a $\frac{1}{2}$ change in the biorthogonal DTOP. (a) and (b) illustrate the biorthogonal Loschmidt rate and biorthogonal DTOP, respectively. The $\frac{1}{2}$ change in the biorthogonal DTOP appears if and only if the prequench phases are in the middle of the phase diagram, with winding numbers of $\frac{1}{2}$ or $0$.}
\label{fig3}
\end{figure}

To show that the half-integer jump of $\nu(t)$ is not a fine-tune result \cite{Ding2020PRBR}, we study different quenching processes by changing $\eta$ but with $\gamma$ fixed. Figures~\ref{fig3}(a) and \ref{fig3}(b) present five typical behaviors of the biorthogonal Loschmidt rate $\text{LR}\left(t \right)$ and biorthogonal DTOP $\nu(t)$, respectively. And more detailed information can be found in the Supplemental Material \cite{SM}. The half-integer jump of $\nu(t)$ can appear alone, periodically, or accompanied by an integer jump, exhibiting rich behavior in a single non-Hermitian Hamiltonian. We also find that the half-integer jump phenomenon occurs if and only if the prequench phases are in the middle of the phase diagram with 0 or 1/2 winding number.

\begin{figure}[ptb]
\begin{center}
\includegraphics[width=8.6cm]{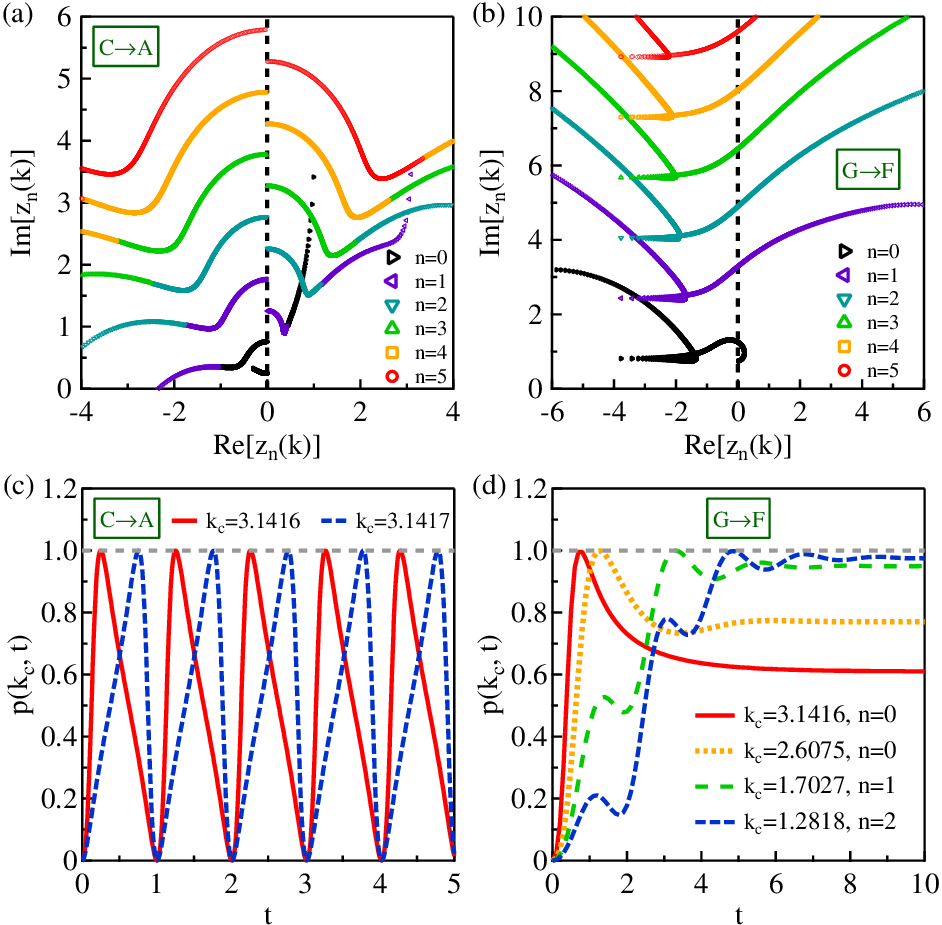}
\end{center}
\caption{(a), (b) The dynamical Fisher zeros in the complex plane $z_n(k)=\text{i}t_n(k)$ with $k$ lies in the first Brillouin zone for two typical quench processes.  (c), (d) The corresponding transition probability $p(k_c,t)$ between $|u_{k_c-}^i(t)\rangle$ and $|u_{k_c+}^i\rangle$ at the critical momentum $k_c$.}
\label{fig4}%
\end{figure}

To better understand the biorthogonal DQPTs, we study the dynamical counterpart of Fisher zeros \cite{Fisher1967RPP,Yang1952PR,Lee1952PR,Andraschko2014PRB,Stauber2017PRE} in the complex time plane $z_n(k)=\text{i}t_n(k)$. The lines $z_n(k)$ cross the imaginary time axis at a critical momentum $k_c$ and yield a critical time $t_{c,n}$, as shown in Figs.~\ref{fig4}(a) and \ref{fig4}(b). The discontinuity of the dynamical Fisher zeros over the imaginary axis is associated with the half-integer jump of $\nu(t)$ \cite{Mondal2022PRB}. Figures~\ref{fig4}(a) and \ref{fig4}(b) also represent two typical types of biorthogonal DQPTs, depending on whether the lines $z_n(k)$ cut the time axis periodically or nonperiodically. In order to further study these two types of biorthogonal DQPTs, we also investigate the transition probability $p(k,t)$ between the time-evolved state $|u_{k-}^i(t)\rangle=\text{e}^{-\text{i}H^f_{k}t}|u_{k-}^i\rangle$ and another initial eigenstate $|u_{k+}^i\rangle$. By expanding $|u_{k-}^i(t)\rangle$ as $|u_{k-}^i(t)\rangle = c_1|u_{k-}^i\rangle +c_2|u_{k+}^i\rangle$, we have $p(k,t) = c_{2}^{\ast}c_{2}/(c_{1}^{\ast}c_{1}+c_{2}^{\ast}c_{2})$ from Eq.~\eqref{projection}. If there exists $p(k,t)=1$, the time-evolved state $|u_{k-}^i(t)\rangle$ is biorthogonal with $|u_{k-}^i\rangle$. Then the biorthogonal Loschmidt echo equals to zero because we can rewrite $\mathcal{L}(t)$ as
\begin{align}
\mathcal{L}(t)=\prod_k\frac{\langle\widetilde{u}_{k-}^i|u_{k-}^i(t)\rangle\langle\widetilde{u}_{k-}^i(t)|u_{k-}^i\rangle}{\langle\widetilde{u}_{k-}^i(t)|u_{k-}^i(t)\rangle}.
\label{another LE}
\end{align}
As shown in Figs.~\ref{fig4}(c) and \ref{fig4}(d), the two types of biorthogonal DQPTs exhibit two distinct behaviors of $p(k,t)$. For the periodic biorthogonal DQPTs, $p(k_c,t)$ oscillate periodically between $0$ and $1$ for fixed critical momenta $k_c$. In this situation, the two-level systems of $k_c$ dominate. Thus the periodicity of biorthogonal DQPTs may be related to the oscillations of the two-level system \cite{Zakrzewski2022arXiv,Damme2023PRR}. On the other hand, $p(k_c,t)$ exhibit very interesting behavior when the biorthogonal DQPTs are no longer periodic. There are many critical momenta $k_c$, and each $k_c$ corresponds to only one $t_{c,n}$. Before $t_{c,n} = - \text{i} z_n(k_c)$, there are $n$ local maximum in $p(k_c,t)$, while for $t \gg t_{c,n}$, $p(k_c,t)$ tends to a fixed value, indicating a steady state in contrast to the oscillation behavior.

In summary, we propose a new theoretical framework to study the biorthogonal DQPTs in non-Hermitian systems based on the biorthogonal quantum mechanics. We reformulate the transition probability between one state $\left\vert \Psi\left(  0\right)  \right\rangle$ and its time-evolved state $\left\vert \Psi\left(  t\right)  \right\rangle$ with the concept of the associated state. Our scheme can handle a general non-Hermitian Hamiltonian with complex eigenvalues, and the normalization factors can be introduced naturally. To demonstrate our approach, we use the non-Hermitian Su-Schrieffer-Heeger model as a concrete example. Comparing with the self-normal cases, we observe a clear 1/2 change in the biorthogonal DTOP. Furthermore, our results show that the periodicity of biorthogonal DQPTs depends on whether the two-level subsystem at the critical momentum oscillates or reaches a steady state. Our work opens up avenues for exploring the rich biorthogonal DQPTs in non-Hermitian systems. An interesting topic for further investigation would be the study of the mechanism behind the 1/2 change in the biorthogonal DTOP.

We would like to acknowledge valuable discussions with S. Chen. This work is supported by the National Natural Science Foundation of China Grants No. 12204075, No. 11974064, No. 12075040, and No. 12347101; the China Postdoctoral Science Foundation Grant No. 2023M730420; the fellowship of Chongqing Postdoctoral Program for Innovative Talents Grant No. CQBX202222; the Natural Science Foundation of Chongqing Grant No. CSTB2023NSCQ-MSX0953; the Chongqing Research Program of Basic Research and Frontier Technology Grants No. cstc2021jcyjmsxmX0081 and No. cstc2020jcyj-msxmX0890; Chongqing Talents: Exceptional Young Talents Project No. cstc2021ycjh-bgzxm0147; and the Fundamental Research Funds for the Central Universities Grants No. 2020CDJQY-Z003, No. 2022CDJJCLK001, and No. 2021CDJQY-007.

\end{document}


\title{Supplementary Materials for ``Biorthogonal Dynamical Quantum Phase Transitions in Non-Hermitian Systems"}
\author{Yecheng Jing}
\affiliation{Department of Physics and Chongqing Key Laboratory for Strongly Coupled Physics, Chongqing University, Chongqing 401331, China}
\author{Jian-Jun Dong}
\email[]{dongjianjun@cqu.edu.cn}
\affiliation{Department of Physics and Chongqing Key Laboratory for Strongly Coupled Physics, Chongqing University, Chongqing 401331, China}
\author{Yu-Yu Zhang}
\email[]{yuyuzh@cqu.edu.cn}
\affiliation{Department of Physics and Chongqing Key Laboratory for Strongly Coupled Physics, Chongqing University, Chongqing 401331, China}
\author{Zi-Xiang Hu}
\email[]{zxhu@cqu.edu.cn}
\affiliation{Department of Physics and Chongqing Key Laboratory for Strongly Coupled Physics, Chongqing University, Chongqing 401331, China}

\begin{abstract}
In this supplementary material, we will show more details to support the discussion in the main text. In Sec. I, we provide a simple example to illustrate the associate state and time evolution problems in non-Hermitian systems. In Sec. II, we show the detail calculations of the biorthogonal Loschmidt echo and biorthogonal dynamical phase. In Sec. III, we present more detailed information of biorthogonal dynamical quantum phase transition.
\end{abstract}

\maketitle

\renewcommand\thesection{\arabic{section}}
\renewcommand\thefigure{S\arabic{figure}}
\renewcommand\theequation{S\arabic{equation}}
\setcounter{figure}{0}
\setcounter{equation}{0}

\subsection{I. A simple example to illustrate the associate state and time evolution problems in non-Hermitian systems}
\label{appA}

In order to illustrate how to use the concept of associate state to solve the time-evolve states in non-Hermitian systems, we take $2\times2$ non-Hermitian matrices containing complex eigenvalues as an example.
For simplicity, we consider the quench process from $H^i$ to $H^f$ with
\begin{align}
H^i=\left(\begin{matrix} 0 & 4+\text{i} \\ 2-\text{i} & 0 \end{matrix} \right),
\end{align}
and
\begin{align}
H^f=\left( \begin{matrix} 0 & -3\text{i} \\
-2+3\text{i} & 0 \end{matrix} \right).
\end{align}
The biorthogonal bases of the initial Hamiltonian $H^i$ are given by
\begin{align}
|\psi_{\pm}\rangle=\frac{1}{\sqrt{2}}
\left( \begin{matrix} \pm\sqrt{\frac{4+\text{i}}{2-\text{i}}} \\1\end{matrix} \right), \; |\widetilde{\psi}_{\pm}\rangle=\frac{1}{\sqrt{2}}
\left(\begin{matrix} \pm\sqrt{\frac{2+\text{i}}{4-\text{i}}} \\ 1 \end{matrix} \right).
\end{align}

Now let us consider the time evolution of $|\psi_+\rangle$ under the final Hamiltonian $H^f$, $|\psi_+(t)\rangle=\text{e}^{-\operatorname*{i}H^f t}|\psi_+\rangle$. When $t=1$, it is given by
\begin{equation}
\left\vert \psi_{+}\left(  t=1\right)  \right\rangle \approx\left( \begin{array}[c]{c}%
-1.132+0.190\operatorname*{i}\\
-0.359-0.927\operatorname*{i}\end{array}\right)  .
\end{equation}
It can be expand as
\begin{equation}
\left\vert \psi_{+}\left(  t=1\right)  \right\rangle =a\left\vert \psi
_{+}\right\rangle +b\left\vert \psi_{-}\right\rangle,
\end{equation}
with
\begin{equation}
a\approx-0.772-0.359\operatorname*{i},\quad b\approx0.264-0.953\operatorname*{i}.
\end{equation}
According Eq.~(2) in the main text, the associated state of $\left\vert \psi_{+}\left(  t=1\right)  \right\rangle$ is given by
\begin{align}
|\widetilde{\psi}_{+}\left(  t=1\right)  \rangle &  =a|\widetilde{\psi}%
_{+}\rangle+b|\widetilde{\psi}_{-}\rangle\nonumber\\
&  \approx\left(
\begin{array}
[c]{c}%
-0.614+0.103\operatorname*{i}\\
-0.359-0.927\operatorname*{i}%
\end{array}
\right)  .
\end{align}
We can calculate the transition probability between the time-evolve state $|\psi_+(t)\rangle$ and another arbitrary state, for instance, $|\phi\rangle=2|\psi_+\rangle+3|\psi_-\rangle$, whose associated state is given by $|\widetilde{\phi}\rangle=2|\widetilde{\psi}_+\rangle+3|\widetilde{\psi}_-\rangle$. By using Eq.~(4) in the main text, i.e.,
\begin{equation}
p\left(  t\right)  =\frac{\langle\widetilde{\psi}_{+}\left(  t\right)
|\phi\rangle\langle\widetilde{\phi}|\psi_{+}\left(  t\right)  \rangle}%
{\langle\widetilde{\psi}_{+}\left(  t\right)  |\psi_{+}\left(  t\right)
\rangle\langle\widetilde{\phi}|\phi\rangle},
\end{equation}
we can obtain $p\left(  t=1\right) \approx 0.603$. It is a real number ranging from 0 to 1, satisfying the requirements of probability interpretation \cite{Brody2014JPA}.

Note that the associate state $|\widetilde{\psi}_{+}\left(  t\right)  \rangle$ no longer follows the Schr\"{o}dinger equation (denoted by subscript $S$), $|\widetilde{\psi}_+(t)\rangle_{S}=\text{e}^{-\operatorname*{i}H^{f^{\dag}} t}|\widetilde{\psi}_+\rangle$. To see this clearly, we can also calculate the transition probability between $|\psi_+(t)\rangle$ and $|\phi\rangle$ as above, but use $|\widetilde{\psi}_+(t)\rangle_{S}$ temporally. When $t=1$, the result is
\begin{equation}
|\widetilde{\psi}_{+}\left(  t=1\right)  \rangle_{S}\approx\left(
\begin{array}
[c]{c}%
-0.967-1.094\operatorname*{i}\\
-1.373+0.411\operatorname*{i}%
\end{array}
\right)  ,
\end{equation}
and
\begin{equation}
p_{S}\left(  t=1\right)  \approx -0.372+1.118\operatorname*{i}.
\end{equation}
It turns out that if one persist on using the Schr\"{o}dinger equation, an unreasonable complex result can be obtained when the system does not have parity-time symmetry \cite{Brody2014JPA,Tang2022EPL}. Therefore, we must use Eq.~(2) in the main text to calculate the associated state when dealing with the time evolution problems in a general non-Hermitian systems.

\subsection{II. Detail calculations of the biorthogonal Loschmidt echo and biorthogonal dynamical phase}
\label{appB}

We can rewrite the biorthogonal Loschmidt echo as
\begin{align}
\mathcal{L}(t)
&=\frac{|\langle\widetilde{\Psi}(0)|\Psi(t)\rangle|^2}{\langle\widetilde{\Psi}(t)|\Psi(t)\rangle\langle\widetilde{\Psi}(0)|\Psi(0)\rangle}\nonumber\\
&=\prod_k\frac{|\langle\widetilde{u}_{k-}^i|u_{k-}^i(t)\rangle|^2}{\langle\widetilde{u}_{k-}^i(t)|u_{k-}^i(t)\rangle\langle\widetilde{u}_{k-}^i|u_{k-}^i\rangle}\nonumber\\
&=\prod_k g_k(t),
\end{align}
where we have introduced
\begin{align}
g_k(t)&=\frac{|\langle\widetilde{u}_{k-}^i|u_{k-}^i(t)\rangle|^2}{\langle\widetilde{u}_{k-}^i(t)|u_{k-}^i(t)\rangle}.
\end{align}
We then calculate the numerator in $g_k(t)$,
\begin{align}
\langle\widetilde{u}_{k-}^i|u_{k-}^i(t)\rangle=&\langle \widetilde{u}_{k-}^i|u_{k-}^f\rangle\langle \widetilde{u}_{k-}^f|\text{e}^{-\text{i}H_k^ft}|u_{k-}^f\rangle\langle \widetilde{u}_{k-}^f|u_{k-}^i\rangle\nonumber\\
&+\langle \widetilde{u}_{k-}^i|u_{k+}^f\rangle\langle \widetilde{u}_{k+}^f|\text{e}^{-\text{i}H_k^ft}|u_{k+}^f\rangle\langle \widetilde{u}_{k+}^f|u_{k-}^i\rangle\nonumber\\
=&\cos(\epsilon_{k}^ft)-\text{i}\sin(\epsilon_{k}^ft)\langle\widetilde{u}_{k-}^i|\frac{H_k^f}{\epsilon_k^f}|u_{k-}^i\rangle.
\label{detail1}
\end{align}
Note that we only inset the identity $\sum_{n} | u_{n}\rangle\langle \widetilde{u}_{n}|=1$ to evaluate the numerator in $g_k(t)$. It is consistent with the result obtained using the associated state method where the biorthogonal bases of $H^i$ are used, indicating we should also use the biorthogonal bases of $H^i$ rather than $H^f$ to calculate the denominator in $g_k(t)$.

\begin{figure}[ptb]
\begin{center}
\includegraphics[width=8.6cm]{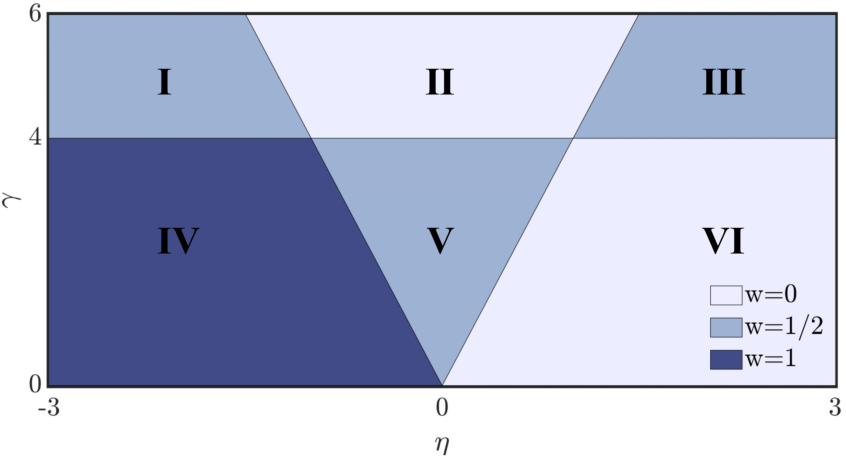}
\end{center}
\caption{The phase diagram of the non-Hermitian Su-Schrieffer-Heeger model. Here, each phase is labeled by a Roman numeral.}
\label{figSM1}
\end{figure}

The expression of biorthogonal dynamical phase can be generalized directly from the definition in Hermitian case \cite{Vajna2015PRB,Ding2020PRBR,Mondal2022PRB},
\begin{align}
\phi_k^{\text{dyn}}(t)
=&-\text{i}\int_{0}^{t}ds\frac{\langle\widetilde{u}_{k-}^i(s)|}{\sqrt{\langle\widetilde{u}_{k-}^i(s)|u_{k-}^i(s)\rangle}}\frac{d}{ds}\frac{|u_{k-}^i(s)\rangle}{\sqrt{\langle\widetilde{u}_{k-}^i(s)|u_{k-}^i(s)\rangle}}\nonumber\\
=&-\text{i}\int_{0}^{t}ds[\frac{\langle\widetilde{u}_{k-}^i(s)|}{\sqrt{\langle\widetilde{u}_{k-}^i(s)|u_{k-}^i(s)\rangle}}\frac{\frac{d}{ds}|u_{k-}^i(s)\rangle}{\sqrt{\langle\widetilde{u}_{k-}^i(s)|u_{k-}^i(s)\rangle}}\nonumber\\
&+\frac{\langle\widetilde{u}_{k-}^i(s)|u_{k-}^i(s)\rangle}{\sqrt{\langle\widetilde{u}_{k-}^i(s)|u_{k-}^i(s)\rangle}}\frac{d}{ds}(\frac{1}{\sqrt{\langle\widetilde{u}_{k-}^i(s)|u_{k-}^i(s)\rangle}})]\nonumber\\
=&-\int_{0}^{t}ds\frac{\langle\widetilde{u}_{k-}^i(s)|H_k^f|u_{k-}^i(s)\rangle}{\langle\widetilde{u}_{k-}^i(s)|u_{k-}^i(s)\rangle}\nonumber\\
&+\frac{\text{i}}{2}\text{ln}[\langle\widetilde{u}_{k-}^i(t)|u_{k-}^i(t)\rangle].   \label{detail2}
\end{align}

\begin{table}[ptb]
\centering
\caption{Comparison of various quench processes, exhibiting rich behavior of biorthogonal DQPTs in a single non-Hermitian Hamiltonian.}
\begin{tabular}{c c c c}
\hline\hline
Quench & Parameters & Fisher zeros profile & DTOP \\   \hline
\uppercase\expandafter{\romannumeral 1} $\leftrightarrow$ \uppercase\expandafter{\romannumeral 2} & (-2, 5)$\leftrightarrow$(0.2, 5) & $\rightarrow:0,1; \ \leftarrow:2$ & $\rightarrow:1; \  \leftarrow:\frac{1}{2}$   \\
\uppercase\expandafter{\romannumeral 1} $\leftrightarrow$ \uppercase\expandafter{\romannumeral 3} & (-2, 5)$\leftrightarrow$(2, 5) & $\leftrightarrow$: 1 & $\leftrightarrow$: 1    \\
\uppercase\expandafter{\romannumeral 2} $\leftrightarrow$ \uppercase\expandafter{\romannumeral 3} & (0.2, 5)$\leftrightarrow$(2, 5) & $\rightarrow:2,3; \ \leftarrow:0$ & $\rightarrow:\frac{1}{2},1; \ \leftarrow:0$    \\
\uppercase\expandafter{\romannumeral 4} $\leftrightarrow$ \uppercase\expandafter{\romannumeral 5} & (-2, 1)$\leftrightarrow$(0.2, 1) & $\rightarrow:1; \ \leftarrow:3,\;4$ & $\rightarrow:1; \ \leftarrow:\frac{1}{2},1$    \\
\uppercase\expandafter{\romannumeral 4} $\leftrightarrow$ \uppercase\expandafter{\romannumeral 6} & (-2, 1)$\leftrightarrow$(2, 1) & $\leftrightarrow$: 2 & $\leftrightarrow$: 1  \\
\uppercase\expandafter{\romannumeral 5} $\leftrightarrow$ \uppercase\expandafter{\romannumeral 6} & (0.2, 1)$\leftrightarrow$(3, 1) & $\rightarrow:2,4; \ \leftarrow:1$ & $\rightarrow:\frac{1}{2},1; \ \leftarrow:1$    \\
\uppercase\expandafter{\romannumeral 1} $\leftrightarrow$ \uppercase\expandafter{\romannumeral 1} & (-2, 5)$\leftrightarrow$(-3, 5) & $\leftrightarrow$: 0,\;1 & $\leftrightarrow$: 1   \\
\uppercase\expandafter{\romannumeral 2} $\leftrightarrow$ \uppercase\expandafter{\romannumeral 2} & (0.2, 5)$\leftrightarrow$(-0.2, 5) & $\leftrightarrow$: 0,\;1 & $\leftrightarrow$: $\frac{1}{2}$  \\
\uppercase\expandafter{\romannumeral 3} $\leftrightarrow$ \uppercase\expandafter{\romannumeral 3} & (2, 5)$\leftrightarrow$(3, 5) & $\leftrightarrow$: 1 & $\leftrightarrow$: 1    \\
\uppercase\expandafter{\romannumeral 4} $\leftrightarrow$ \uppercase\expandafter{\romannumeral 4} & (-2, 1)$\leftrightarrow$(-1, 1) &$\leftrightarrow$: 0,\;2 & $\leftrightarrow$: 1\\
\uppercase\expandafter{\romannumeral 5} $\leftrightarrow$ \uppercase\expandafter{\romannumeral 5} & (0.2, 1)$\leftrightarrow$(-0.2, 1) & $\leftrightarrow$: 1,\;2 & $\leftrightarrow$: $\frac{1}{2}$,\;1\\
\uppercase\expandafter{\romannumeral 6} $\leftrightarrow$ \uppercase\expandafter{\romannumeral 6} & (0.5, 1)$\leftrightarrow$(5, 1) & $\leftrightarrow$: 0,\;2 & $\leftrightarrow$: 1 \\ \hline\hline
\end{tabular}
\label{table}
\end{table}

\begin{figure}[pb]
\begin{center}
\includegraphics[width=8.6cm]{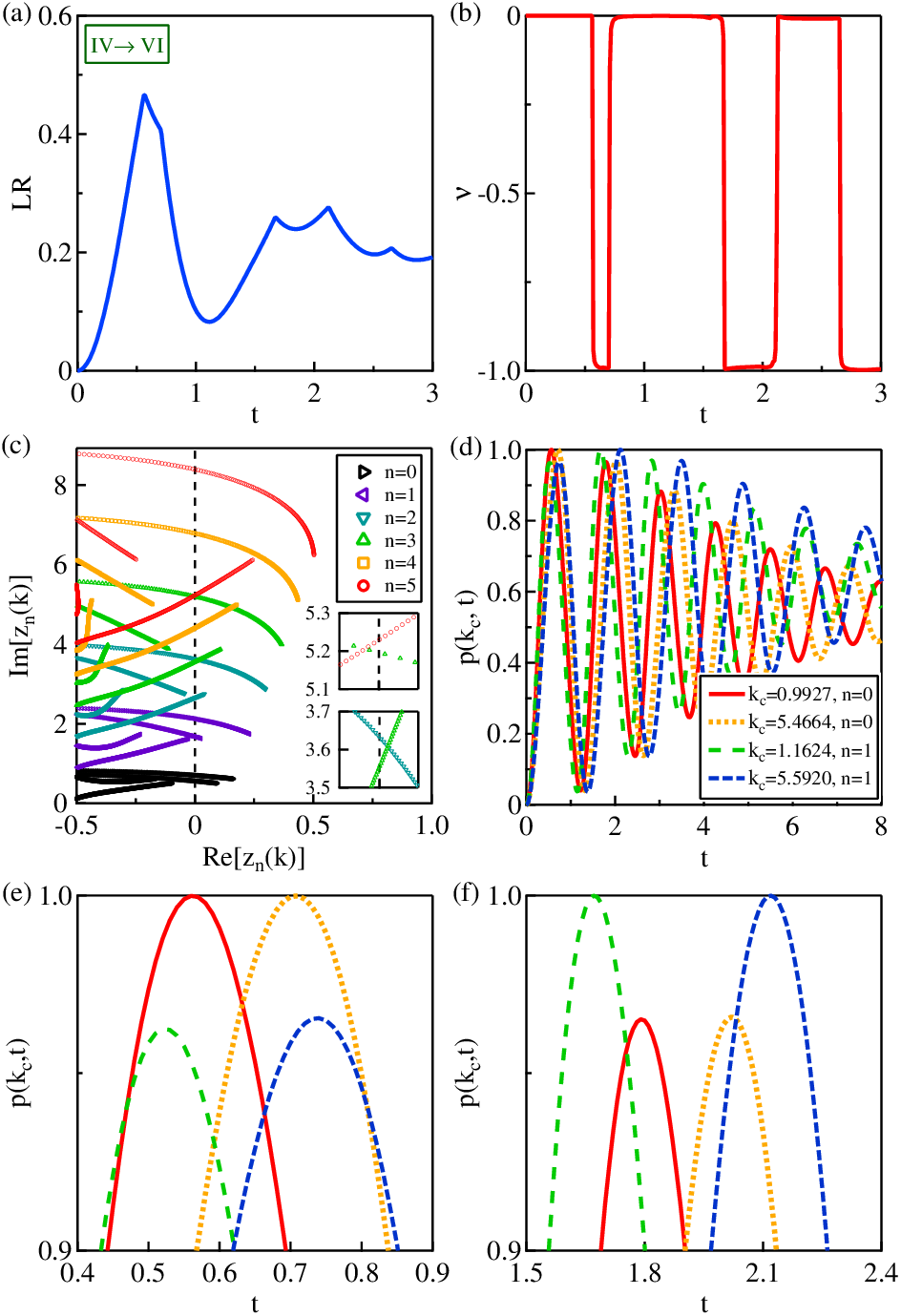}
\end{center}
\caption{(a) The biorthogonal Loschmidt rate, (b) biorthogonal DTOP, (c) dynamical Fisher zeros, and (d) the corresponding two-level transition probabilities $p(k_c,t)$ for a typical quench process in Table {\ref{table}} (from IV to VI). Figures (e) and (f) are the enlarged plots of (d) to highlight the four critical points in (d).}
\label{figSM2}
\end{figure}

\subsection{III. More detailed information of biorthogonal DQPTs}
\label{appC}

There are many types of quenching processes due to the rich phase diagram of the non-Hermitian Su-Schrieffer-Heeger model. For clarity, we label each phase of the non-Hermitian Su-Schrieffer-Heeger model with a Roman numeral, as shown in Fig.~{\ref{figSM1}}. We have shown five typical types of quenching processes in the main text, which contain the most important information. As a supplement, we summarize various quenching processes and the corresponding dynamical Fisher zeros and biorthogonal DTOPs in Table {\ref{table}}. The first column specifies the type of quench between phases, where the direction of the arrow ``$\rightarrow$" and ``$\leftarrow$" denotes the direction of quench. The second column provides concrete parameters $(\eta,\ \gamma)$ of such a quench,
The ``Fisher zeros profile" column indicates the possible number of dynamical Fisher zeros of a branch $z_n(k)$ for a fixed $n$. As mentioned in the main text, a biorthogonal DQPT corresponds to a dynamical Fisher zero \cite{Heyl2013PRL,Andraschko2014PRB,Stauber2017PRE}. Thus the number of dynamical Fisher zeros corresponds to the number of biorthogonal DQPTs in this $z_n(k)$ branch. The final column indicates the possible value of the change in biorthogonal DTOP that may occur during this quench process.

\begin{figure}[ptb]
\begin{center}
\includegraphics[width=8.6cm]{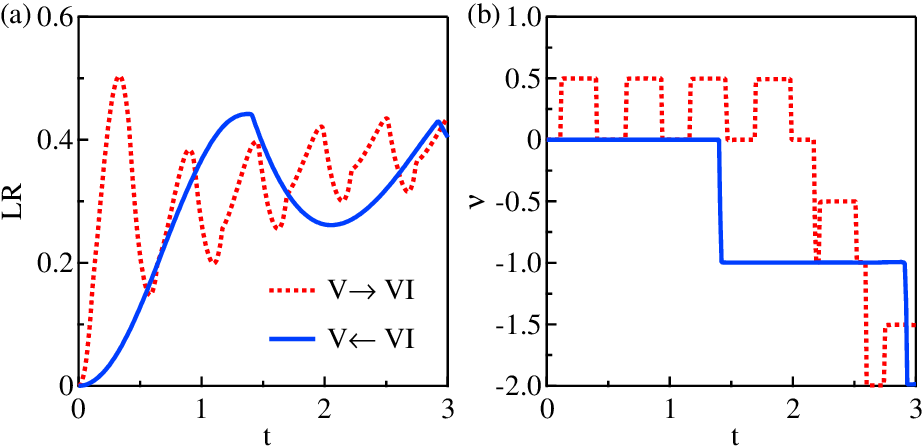}
\end{center}
\caption{(a) The biorthogonal Loschmidt rate and (b) the biorthogonal DTOP for two typical quenching processes (between V and VI) with different directions.}
\label{figSM3}
\end{figure}

\begin{figure}[ptb]
\begin{center}
\includegraphics[width=8.6cm]{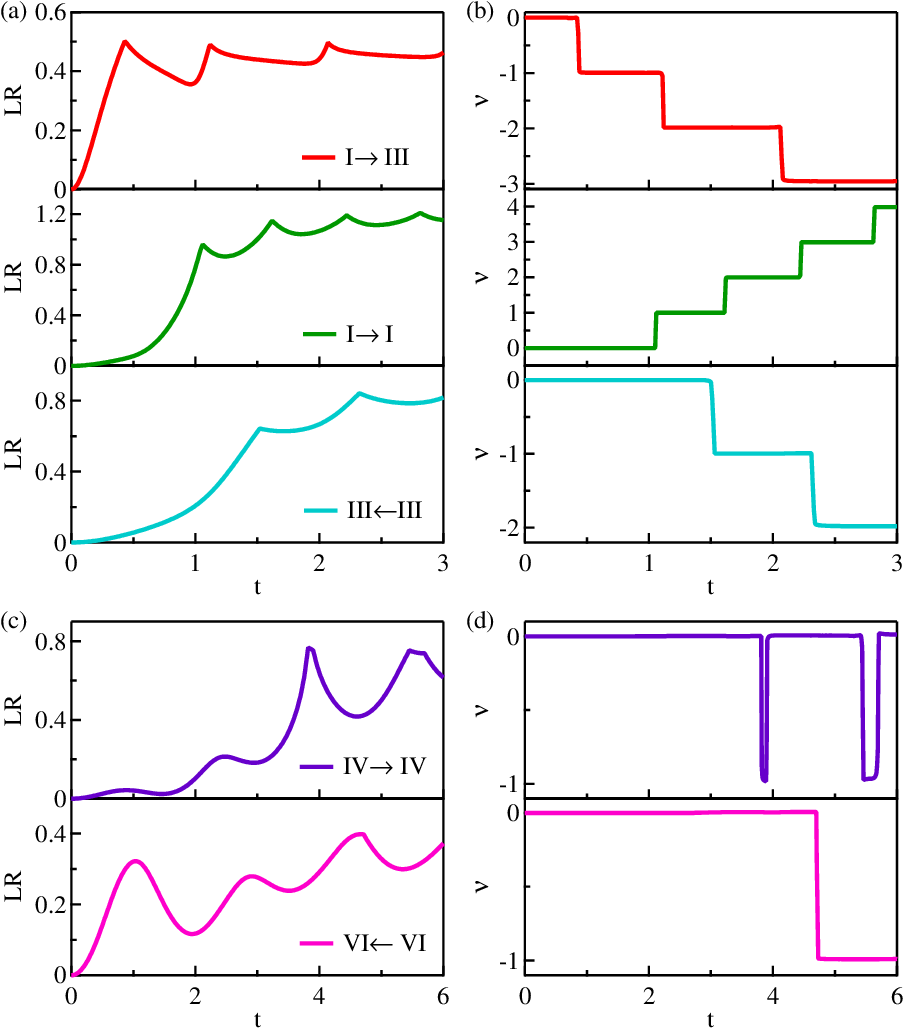}
\end{center}
\caption{The other five types of quench processes in Table {\ref{table}}. The last four quenching processes within the same phase are specified by their direction from top to bottom as $\rightarrow$, \; $\leftarrow$,\;  $\rightarrow$,\; $\leftarrow$, and the corresponding parameter are presented in Table {\ref{table}} with the same arrow. Figures (a) and (c) illustrate the biorthogonal Loschmidt rate, while (b) and (d) show the corresponding biorthogonal DTOPs.}
\label{figSM4}
\end{figure}

To provide a clearer explanation of the information in Table {\ref{table}}, we take a quench process from phase IV to VI as a concrete example. The biorthogonal Loschmidt rate, biorthogonal DTOP, dynamical Fisher zeros and the corresponding two-level transition probabilities are calculated and presented in Fig.~{\ref{figSM2}}. As shown in Fig.~{\ref{figSM2}}(a), the non-analytical behaviors of the biorthogonal Loschmidt rate indicate the occurrence of biorthogonal DQPTs. The jumps in the biorthogonal DTOP value in Fig.~{\ref{figSM2}}(b) highlight the biorthogonal DQPTs. Since the change in the biorthogonal DTOP value can only be one for this quench process, we fill in the number 1 in the ``DTOP" column of the sixth row in Table {\ref{table}}. In Fig.~{\ref{figSM2}}(c), the dynamical Fisher zeros of each branch $z_n(k)$ for a fixed $n$ (distinguished by different colors) are two, which has been recorded in the ``Fisher zeros profile" column. The insets of Fig.~{\ref{figSM2}}(c) show the subtle differences among various critical time. Finally, Fig.~{\ref{figSM2}}(d) shows the corresponding two-level transition probabilities as we did in the main text. Figures~{\ref{figSM2}}(e) and {\ref{figSM2}}(f) are the enlarged plots of {\ref{figSM2}}(d) which are used to display the four DQPTs more clearly since they are indistinguishable in Fig.~{\ref{figSM2}}(d).

Another information contained in Table {\ref{table}} is that the direction of the quenching processes can change the results dramatically. As shown in Fig.~{\ref{figSM3}}, the change of biorthogonal DTOP can be 1/2 or 1 when the quenching process is ``V $\rightarrow$ VI". We use ``$\rightarrow:\frac{1}{2},1$" to represent this and fill it into the corresponding position of Table {\ref{table}}. However, the change of biorthogonal DTOP is always 1 when the quenching process is reversed, i.e., from phase VI to phase V, ``V $\leftarrow$ VI". Correspondingly, we use ``$\leftarrow: 1$" to represent it as indicated in Table {\ref{table}}. It can also be seen from Table {\ref{table}} that when the quenching parameters are in the same phase, the change of biorthogonal DTOP in the two directions are same with each other. We use double-headed arrow ``$\leftrightarrow$" to denote this. Besides quenching in the same phases, there are other cases where the change of biorthogonal DTOP is independent of the quenching direction, such as I $ \leftrightarrow $ III and IV $ \leftrightarrow $ VI.

In order to make other quenching processes in Table {\ref{table}} more visible, we also present the corresponding biorthogonal DQPTs in Fig.~{\ref{figSM4}}. Figures~{\ref{figSM4}}(a) and {\ref{figSM4}}(c) illustrate the biorthogonal Loschmidt rate, while Figs.~{\ref{figSM4}}(b) and {\ref{figSM4}}(d) show the corresponding biorthogonal DTOPs. As discussed in the main text, the half-jump phenomenon occurs if and only if the prequench phases are the phases II or V. The prequench parameters we present here are not in these two phases, so the change of DTOP are all 1.